\documentclass[prd,12pt,tightenlines,nofootinbib]{revtex4}
\usepackage[english]{babel}
\usepackage{graphicx}
\usepackage{psfrag}
\usepackage{amsmath}
\usepackage{amssymb}
\usepackage{bbm}
\usepackage{slashed}
\usepackage{verbatim}
\usepackage[hypertex]{hyperref}

\newcommand{\be}{\begin{equation}}
\newcommand{\ee}{\end{equation}}
\newcommand{\bea}{\begin{eqnarray}}
\newcommand{\eea}{\end{eqnarray}}
\newcommand{\bean}{\begin{eqnarray*}}
\newcommand{\eean}{\end{eqnarray*}}

\renewcommand{\b}{\langle}
\newcommand{\ket}{\rangle}

\newcommand{\e}{{\rm e}}

\renewcommand{\d}{{\rm d}}

\newcommand{\ssst}{\scriptscriptstyle}

\newcommand{\eq}[1]{(\ref{#1})}

\newcommand{\fig}[1]{Fig.\ \ref{#1}}

\newcommand{\qed}{\nobreak \ifvmode \relax \else
      \ifdim\lastskip< 1 em \hskip-\lastskip
      \hskip1.0em plus0em minus0.5em \fi \nobreak
      \vrule height0.75em width0.75em depth0 em\fi}

\newcommand{\xib}{\overline{\xib}}

\begin{document}

\title{Space as a low--temperature regime of graphs}

\author{Florian Conrady}
\email{fconrady@perimeterinstitute.ca}
\affiliation{Perimeter Institute for Theoretical Physics, Waterloo, Ontario, Canada}

\begin{abstract}
I define a statistical model of graphs in which
2--dimensional spaces arise at low temperature.
The configurations are given by graphs with a fixed number
of edges and the Hamiltonian is a simple, local function of
the graphs. Simulations show that there is a
transition between a low--temperature regime in which
the graphs form triangulations of 2--dimensional surfaces
and a high--temperature regime, where the surfaces disappear.
I use data for the specific heat and other observables to discuss
whether this is a phase transition.
The surface states are analyzed with regard to topology and defects.
\end{abstract}

\maketitle

\section{Introduction}
\label{introduction}

Spacetime represents one of the remaining frontiers in theoretical physics.
We have successful theories of fields \textit{in} spacetime and of the geometry \textit{of} spacetime.
But there is no established theory of spacetime itself, a theory that would explain why we live in
a spacetime and not in something completely different.

There are approaches that go in this direction, notably string theory, matrix models \cite{Ishibashietal}--\cite{SmolinMatrixuniversality},
twistor theory \cite{HuggettTod,PenroseRindlerVol2},
causal dynamical triangulations \cite{CDTQuestforQG,CDTArtof}, Regge calculus \cite{HamberReview},
loop quantum gravity \cite{Thiemannbook,Rovellibook,Perezreview}
and quantum graphity \cite{KonopkaMarkopoulouSmolin}--\cite{CaravelliMarkopoulou}.
Some of these start from objects that are radically different from spacetime
and one hopes that spacetime arises from the dynamics of these objects.
In others, spacetime is assumed from the beginning and only the topology and geometry is dynamical.

In this article, I am concerned with the former type of approach, i.e.\ the attempt to
obtain space from a theory, where there is no space to start with.
I define a simple statistical model of graphs that is designed to have 2--dimensional
triangulations as ground states.
The setup is close in spirit to earlier models of quantum graphity explored by
Konopka, Markopoulou, Smolin and others \cite{KonopkaMarkopoulouSmolin}--\cite{CaravelliMarkopoulou}.
I present data from Monte Carlo simulations\footnote{For an introduction to Monte Carlo methods,
see e.g.\ \cite{NewmanBarkema}.} of graphs with 45, 90 and 180 edges.
The results show that the system settles into 2d triangulations at low temperatures.
Above a transition temperature the surface states disappear and more general graphs
dominate.
As far as I know, this is the first time that a model of graphs succeeds in reaching
manifold--like states without help from additional ad hoc constraints.
The temperature dependence of specific heat and other observables
exhibits similar features as in phase transitions of more conventional systems.
The transition temperature, however, drifts as the size of the system is changed,
leaving both a finite-- and zero--temperature phase transition as a
possible scenario for the infinite--size limit.

In a second set of simulations, I investigated the surface properties
in more detail with regard to topology, defects and curvature.
At very low temperature the surfaces are typically connected and
non--orientable, and carry a number of conical singularities.

In the space of graphs the surface states form a very complex subset.
Like in other systems with a complex energy landscape
(see e.g.\ \cite{Young,HarmannRicciTersenghi,HartmannFindingLowtemperaturestates}),
this makes it difficult to ensure equilibrium at low temperature.
The simulation may get trapped in energy valleys and sample
no longer according to the Boltzmann distribution.
To check whether the simulation is equilibrated,
I compare the original simulation
with an alternate simulation schedule that
involves frequent restarts and annealing.

Although triangulations play a role in this model, the setup is quite
different from causal dynamical triangualations
and Regge calculus.
In these theories, each individual configuration is a space
in the form of a triangulation, and one finds different regimes, where
the effective geometry is smooth, crumbled or polymer--like, for example.
In our case, the configurations are graphs and the regimes are distinguished
by the existence or absence of a space.

For testing and inspecting the simulation
I used the Ubigraph software for visualizing dynamic graphs \cite{Ubigraph}.
Videos of the simulation can be viewed at this link \cite{simulation}.

The paper is organized as follows.
In section \ref{themodel} I define the model and discuss the expected ground states
and its defects. I also define an observable used to measure the ``superficiality''
of graphs.
Section \ref{MonteCarlosimulation} describes the details of the algorithm---the
Monte Carlo moves and the trial probabilities needed for detailed balance.
This is followed by sec.\ \ref{resultsofthesimulation}, where I report the results of the simulation.
I discuss evidence for and against a phase transition and the graphs' properties at low
temperature.
The question of low temperature equilibration is scrutinized in
sec.\ \ref{subtletiesatlowtemperature}.
In the final section, I give a summary and discuss some of the conceptual
aspects of the model.

\section{The model}
\label{themodel}

\subsection{Definition}

The model is a statistical model of graphs.
Configurations are given by undirected, labelled graphs $\Gamma$ with a fixed number $N_e$ of edges.
The edges of the graphs are labelled by numbers 1,2,\ldots, $N_e$.
Automorphic graphs are distinguished if they differ in their labelling of edges.
Vertices have to be contained in at least one edge.
There can be at most one edge between two vertices, i.e.\ double, triple etc.\
edges are not allowed. Moreover, edges from a vertex to itself are prohibited.

\renewcommand{\arraystretch}{1.7}
Vertices and edges of graphs are denoted by $v$ and $e$ respectively.
When $n + 1$ vertices form a complete subgraph (i.e.\ when every vertex of the
subgraph is connected with each of its other vertices), I call this subgraph
an $n$--simplex. Thus, vertices and edges define 0-- and 1--simplices.
2-- and 3--simplices will be referred to as triangles $t$ and tetrahedra $\tau$
respectively.
For the definition of the Hamiltonian the notion of \textit{valence} is important.
An $n$--simplex $\Delta$ is said to have valence $V_\Delta$, if it is contained
in $V_\Delta$ $(n + 1)$--simplices.

The Hamiltonian is specified in such a way that it favors graphs corresponding to a
2--dimensional triangulation. That is, graphs are energetically preferred when
its subgraphs form simplices of a 2--dimensional simplicial complex.
The Hamiltonian depends on the graph $\Gamma$ and
is a sum of four types of terms that are associated to vertices,
edges, triangles and tetrahedra:
\be
H_\Gamma = \sum_v H_v + \sum_e H_e + \sum_t H_t + \sum_\tau H_\tau
\label{Hamiltonian}
\ee
The sums extend over all vertices $v$, edges $e$, triangles $t$ and tetrahedra $\tau$
contained in $\Gamma$.
The individual terms are defined as follows:
\be
H_v = c_v \left|V_v - 6\right|\,,\qquad
H_e =
\left\{
\parbox{2.7cm}{
$\begin{array}{rl}
0\,, & V_e \le 2\,, \\
c_e V_e^2\,, & V_e > 2\,,
\end{array}$
}
\right.\,,\qquad
H_t = -c_t\,,\qquad
H_\tau = c_\tau\,.
\ee
The coefficients $c_v$, $c_e$, $c_t$ and $c_\tau$ are positive constants
and their specific values are taken to be
\be
c_v = 1\,,\qquad c_e = 10\,,\qquad c_t = 7\,,\qquad c_\tau = 500\,.
\ee
The most important terms are $H_e$ and $H_t$.
The triangle term $H_t$ favors the creation of triangles, since every triangle lowers the energy by $c_t$.
The edge term $H_e$ penalizes edges with a valence $V_e$ greater than 2. It thus suppresses the
branching of triangles, which would be incompatible with a 2d triangulation.
The tetrahedral term $H_\tau$ suppresses the formation of tetrahedra and allows us to exclude surfaces
that consist merely of disconnected tetrahedra.
The vertex term $H_v$ was introduced in order to favor smooth, low--curvature triangulations over rough,
and crumbled ones\footnote{As such a triangulation defines only a topological manifold, but if we think of each edge as
having the length 1, it becomes a manifold with a metric, i.e.\ a manifold with a notion of distance and curvature.}.
The curvature at a vertex $v$ is determined by its valence $V_v$, and in 2 dimensions a valence of 6 corresponds to flatness.

The partition function is defined by
\be
Z = \sum_{\Gamma} \e^{-\beta H_\Gamma}\,,
\ee
where $\beta$ is the inverse of the temperature $T$. The sum includes all graphs with $N_e$ edges
that are admissible according to the rules stated previously.
The average of an observable $O_\Gamma$ is given by
\be
\b O\ket = \frac{1}{Z} \sum_{\Gamma} O_\Gamma\, \e^{-\beta H_\Gamma}\,.
\ee

\subsection{Surfaces, defects and their observables}

The above choice of the Hamiltonian makes it likely that triangulations of surfaces
are among the ground states of the system. Surfaces, in the strict sense, need not be the only minima of energy, however,
since there can occur defects to which the Hamiltonian is not sensitive.

\begin{figure}
\begin{center}
\includegraphics[height=2.5cm]{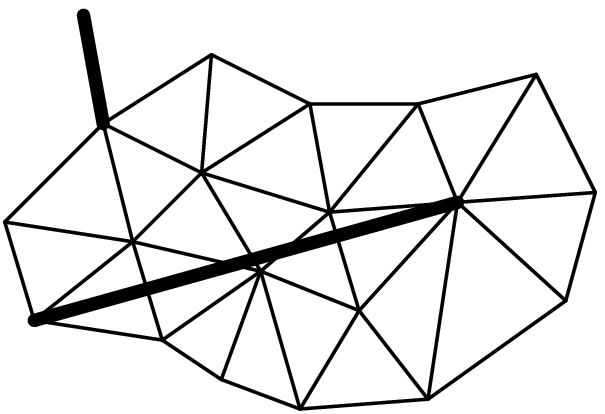}
\hspace{2cm}
\includegraphics[height=2.5cm]{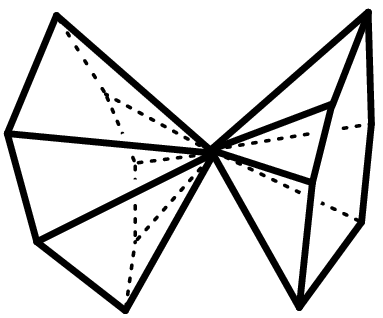}
\end{center}
\caption{Defects: edges with valence $V_e = 0$ and conical singularities.}
\label{defects}
\end{figure}

Through the identification of complete subgraphs with simplices, each admissible graph $\Gamma$
defines a simplicial complex.
A simplicial complex is a triangulation of a surface (with boundary) if and only if\footnote{see e.g.\ \cite{AhlforsSario}, sec.\ I.4.22}
\begin{enumerate}
\item[(a)] all edges have valence $V_e = 1$ or $V_e = 2$, and
\item[(b)] at every vertex edges and triangles form a sequence $e_1\, t_1\, e_2\, t_2\, e_3 \ldots e_n\, t_n\, e_{n+1}$,
where $e_i, e_{i+1}\subset t_i$ (and $e_1 = e_{n+1}$ or $e_1\neq e_{n+1}$).
\end{enumerate}
The term $H_e$ in the Hamiltonian \eq{Hamiltonian} does not distinguish between valence $V_e = 1,2$ and $V_e = 0$,
so there could be minima that consist of surfaces dressed with additional edges outside of any triangle (see \fig{defects}).
The term $H_t$ tends to eliminate these edges in order to maximize the number of triangles for a given
total edge number $N_e$. Nevertheless there could remain edges as ``odd men out'',
since their inclusion would require a restructuring of the triangulation.
The analysis is further complicated by the term $H_v$, which is sensitive to the valence of vertices.
Finally, even when condition (a) is fulfilled, the simplicial complex may not be a surface,
because condition (b) is violated at one or several vertices.
Such vertices correspond to so--called conical singularities, where two surfaces are joined in a single point (see \fig{defects}).

For these reasons, it is difficult to state precisely which graphs are the minima of energy.
The previous arguments suggest that the ground state is given by surfaces with defects
in the form of conical singularities and edges with valence $V_e = 0$.

The purpose of the present work is to determine by simulation whether this intuitive picture
is correct and if the system does indeed settle into surface--like states at low temperatures.
Moreover, this should allow one to see how the system departs from these states as the temperature
is increased. To quantify the presence (or absence) of a surface, I use an observable
$M$ that measures how close the graph is to being a triangulation of a 2d  manifold:
\be
M \equiv \sum_e m_e\,,\qquad
m_e \equiv
\left\{
\parbox{2.7cm}{
$\begin{array}{rl}
0\,, & V_e = 0\,, \\
1/2\,, & V_e = 1\,, \\
1\,, & V_e = 2\,, \\
0\,, & V_e > 2\,.
\end{array}$
}
\right.
\label{observableM}
\ee
Edges with valence $V_e = 2$ contribute with 1 to $M$, boundary edges add 1/2 and any other edge counts as zero.
When comparing systems of different size $N_e$, it is convenient to use the ``intensive'' quantity
\be
m \equiv \frac{1}{N_e} M\,.
\ee
A graph with $m = 1$ is, up to conical singularities, a perfect surface without boundary.
As $m$ is not sensitive to conical singularities, the number $N_{\textrm{con}}$ of such singularities
is another observable that will be measured in the simulation.
By analogy to the susceptibility of a magnet, I will also define and measure a ``susceptibility'' related to $M$:
\be
\chi \equiv \frac{\beta}{N_e}\left(\b M^2\ket - \b M\ket^2\right) = \beta N_e \left(\b m^2\ket - \b m\ket^2\right)
\label{definitionsuseceptibility}
\ee

\section{Monte Carlo simulation}
\label{MonteCarlosimulation}

I simulated the model by the Monte Carlo method using a Metropolis algorithm.
In setting up this simulation the main challenge was to find suitable moves
that allow one to explore the states of the system efficiently at low temperatures.
One can adopt a very simple set of moves that displace edges randomly from one location to another,
and such moves would be clearly ergodic.
It is far from clear, however, that these moves are also powerful
enough to sample all the relevant configurations within a limited simulation time.
The surface configurations deemed to be important at low temperature constitute only a tiny
fraction of the full space of graphs. It is doubtful that a mere hit and miss method
can reach these very special graphs.
Initial simulations showed, in fact, that such minimal moves are not sufficient for
finding these configurations.

To resolve this problem I extended this set of moves by more elaborate moves that assist
the system in building low--energy configurations.
The resulting algorithm succeeds in sampling the surface graphs\footnote{It is another question
whether this sampling is also unbiased, i.e.\ representative of the low--temperature equilibrium.
This will be discussed in section \ref{subtletiesatlowtemperature}.}.

In the next two subsections, I describe the moves of this algorithm
and also the trial probabilities that are needed to achieve detailed balance.

\subsection{Moves}

Since the number $N_e$ of edges is conserved, each Monte Carlo move consists of a succession of two more basic moves.
The first move removes $m$ edges from the graph and the second one adds again $m$ edges.

\begin{figure}
\begin{center}
\includegraphics[height=6cm]{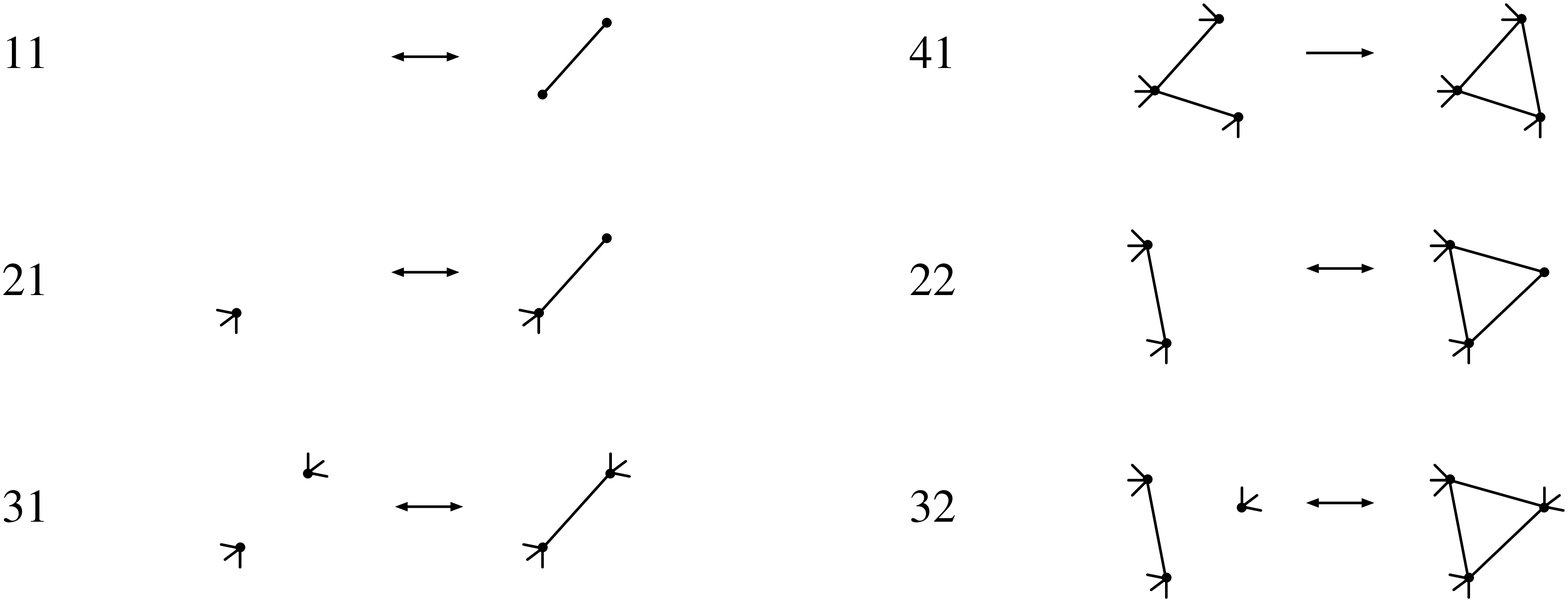}
\end{center}
\caption{List of moves from which Monte Carlo moves are built.}
\label{moves}
\end{figure}

I found it convenient to categorize these moves into 6 types, labelled by a pair of numbers $n_m$.
The number $m$ denotes the change in the number of edges, while $n$ labels the specific way in which the addition
(or removal) of edges happens (see \fig{moves}).

Roughly speaking, $n_m$, $m > 0$, stands for a move that creates a new simplex by adding $m$ edges.
If $n = 1$, this is done by adding a new simplex that is disconncted from the rest of the graph.
If $n = 2$, $m$ edges are ``glued'' onto an existing $(m-1)$--simplex $\Delta_{m-1}$ to form an $m$--simplex $\Delta_m$.
If $n = 3$, the $m$--simplex $\Delta_m$ is built by ``glueing'' $m$ edges to an existing $(m-1)$--simplex $\Delta_{m-1}$
and a vertex.

\psfrag{11}{$1_{\pm 1}$}
\psfrag{21}{$2_{\pm 1}$}
\psfrag{31}{$3_{\pm 1}$}
\psfrag{41}{$4_1$}
\psfrag{22}{$2_{\pm 2}$}
\psfrag{32}{$3_{\pm 2}$}

More concretely, this means that the move $1_1$ adds an edge whose vertices have both valence 1.
The move $2_1$ adds an edge such that one vertex of the edge has valence 1 and the other vertex has valence greater than 1.
The $3_1$ move adds an edge between two existing vertices.

I also introduce a fourth move, called $4_1$.
It adds an edge between two vertices that are linked to a common vertex, but are not linked directly.
These moves are redundant in the sense that they are equivalent to a subset of the $3_1$ moves.
The $4_1$ moves are nevertheless important, since the explicit proposal of such moves makes the algorithm far more efficient
in finding minimum energy configuration. They appear to facilitate the growth of 2--simplices and surfaces.

In the case $m = 2$, we have, in analogy to $2_1$, the $2_2$ move which ``glues'' two edges $a$ and $b$ to an existing edge $c$
such that $a$, $b$, $c$ form a 2--simplex. The vertex shared by $a$ and $b$ has valence 2.
The move $3_2$ does the same as $2_2$ except that the vertex common to $a$ and $b$ has valence greater than 2 (i.e.\
one end of $a$ and $b$ is ``glued'' to an existing vertex of the graph).
Like the $4_1$ move, the $2_1$ and $2_2$ move are not needed for ergodicity,
but they enhance the algorithm's efficiency at low temperature.

In principle, one could also define a $1_2$ move that adds a disconnected 2--simplex.
However, I found it easier to omit this move in the algorithm.
Since the edge number is conserved, $1_2$ would have to be accompanied by
two moves of $m = -1$, which would require additional program code.

The moves $n_m$ with negative $m$ are defined as the inverses of those with $m > 0$.
Altogether we have the following set of basic moves\footnote{We do not need $4_{-1}$, since 2--simplices are easily
destructed by $3_{-1}$.}:
\[
\left\{1_{\pm 1}, 2_{\pm 1}, 3_{\pm 1}, 4_1, 2_{\pm 2},3_{\pm 2}\right\}
\]
Each of the Monte Carlo moves is given by a sequence of moves $(n_{-m},n'_m)$, $m > 0$, where $n_{-m}$
and $n'_m$ are taken from the previous set.

\subsection{Trial probabilities and detailed balance}

Next I describe the way Monte Carlo moves are proposed.
The resulting proposal (or trial) probabilities are generally not symmetric
w.r.t.\ inversion of the move. Therefore, to ensure detailed balance,
we need to include trial probabilities when determining the acceptance probability.

The trial probability for the Monte Carlo move $(n_{-m},n'_m)$ is given by
\be
P_t(n_{-m},n'_m) = P_t(n_{-m}) P_t(n'_m)\,,
\ee
where $P_t(n_{-m})$ and $P_t(n'_m)$ are the trial probabilities for the
subtraction and subsequent addition of edges.

For the subtraction move $n_{-m}$, $m > 0$, the selection proceeds as follows.
There is a probability $P_m$ of choosing $m = 1$ or $m = 2$ that is weighted in
a specific manner by the number of existing simplices.
Then, we have a probability $P_{\Delta_m}$ for choosing the $m$--simplex $\Delta_m$ on which
the move operates. This simplex is chosen randomly from the set of $m$--simplices, so
\be
P_{\Delta_m} = \frac{1}{N_{\Delta_m}}\,.
\ee
Finally, one has to select the type $n$ of the move. The probability $P_n$ for this is
1 in all cases except for $3_{-2}$, where it is $P_3 = 1/3$, since there
are three ways of removing two edges from a given 2--simplex.
The overall trial probability for the subtraction move $n_{-m}$ is
\be
P_t(n_{-m}) = P_m P_{\Delta_m} P_n\,.
\ee
In the case of the additive move, we do not have to choose $m$, since it is fixed
by the previous removal of edges.
The choice of type $n'$ is based on a certain weighting $P_{n'}$ whose details we
omit.
The non--trivial part is the probability $P_{\textrm{rec}}(n'_m)$ for selecting
the simplices that serve as ``receptors'' for the edges to be added.
For moves of type 1, $P_{\textrm{rec}}(1_m) = 1$. For $n' = 2$, we have
\be
P_{\textrm{rec}}(2_m) = P_{\Delta_{m-1}} = \frac{1}{N_{\Delta_{m-1}}}\,,
\ee
where $P_{\Delta_{m-1}}$ is the likelihood of picking one of the $(m-1)$--simplices
at random. For $n' = 3$, one has to choose both an $(m-1)$--simplex and a vertex,
therefore
\be
P_{\textrm{rec}}(3_m) = P_{\Delta_{m-1}} P_v = \frac{1}{N_{\Delta_{m-1}} N_v}\,.
\ee
The total trial probability for $n'_m$ is
\be
P_t(n'_m) = P_{n'} P_{\textrm{rec}}(n'_m)\,.
\ee
An exception to this rule occurs when we deal with a move that is both $3_1$ and $4_1$.
In this case, we have
\be
P_t(3_1) \equiv P_t(4_1) = P_3 P_{\textrm{rec}}(3_1) + P_4 P_{\textrm{rec}}(4_1)\,,
\ee
where $P_{\textrm{rec}}(3_1)$ is given as before and $P_{\textrm{rec}}(4_1)$ is
the likelihood of proposing the receptors of the $4_1$ move. The latter are determined by
first choosing a vertex $v_c$ that I call the ``center'' vertex
and then vertices $v_1$ and $v_2$ that are linked to
$v_c$, but not linked to each other (see \fig{moves}).
Thus, the likelihood for a given pair $(v_1,v_2)$ can be written as
\be
P_{\textrm{rec}}(4_1) = \frac{1}{N_v} \sum_{v_c} \frac{1}{N_{v_c}}\,.
\ee
Here, the sum extends over all center vertices $v_c$ compatible with the pair $(v_1,v_2)$,
while $N_{v_c}$ is the number of \textit{all} pairs $(v_1,v_2)$ for which $v_c$ is a center
vertex.

Once the trial probabilities are determined, we can evaluate the acceptance probability $P_a(n_{-m},n'_m)$
via
\be
P_a(n_{-m},n'_m) = \min\left\{ 1\;,\; \frac{P_t(n'_{-m},n_m)}{P_t(n_{-m},n'_m)}\, \e^{-(E'-E)/T} \right\}\,.
\ee
$E$ and $E'$ denote the energy before and after the move.

\section{Results of the simulation}
\label{resultsofthesimulation}

In this section, I present the results of the simulation.
Measurements were done for three sizes of the system, $N_e = 45$, $90$ and $180$,
and for a range of temperatures from $T = 0.5$ to $T = 6.75$.
I begin with the estimates for equilibration times (the next subsection).
Then, in subsection \ref{evidenceofaphasetransition},
I describe the results for the two main observables, the energy $E$ and the observable $M$ (defined in eq.\ \eq{observableM}),
and the quantities derived from them, the specific heat per edge $c$ and the susceptibility $\chi$ associated to $M$
(see eq.\ \eq{definitionsuseceptibility}).
In subsection \ref{surfaceproperties}, the low--temperature regime
is investigated in more detail with regard to topology, curvature and defects.
In section \ref{subtletiesatlowtemperature}, I will discuss the issue of equilibrium at low temperature
and caveats entailed by it.

\subsection{Equilibration times}
\label{equilibrationtimes}

\begin{figure}
\begin{center}
\includegraphics{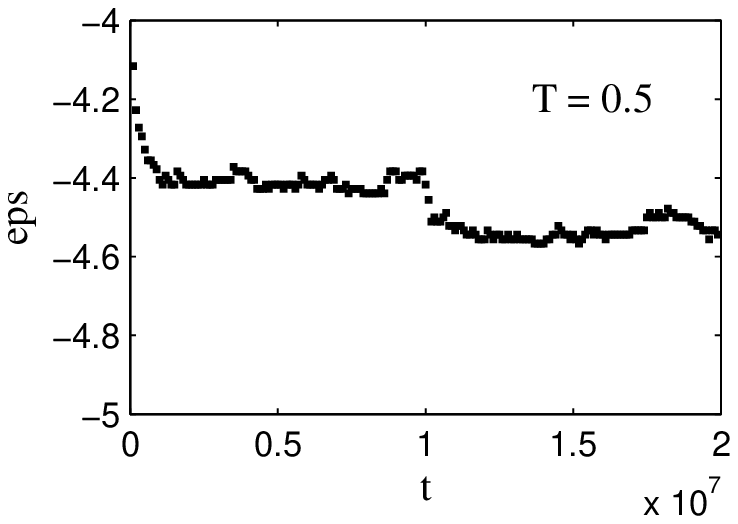}
\hspace{0.1cm}
\includegraphics{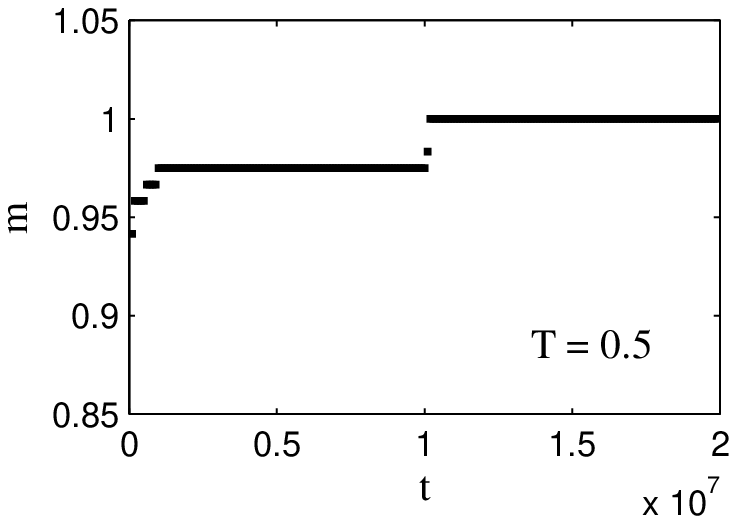}

\includegraphics{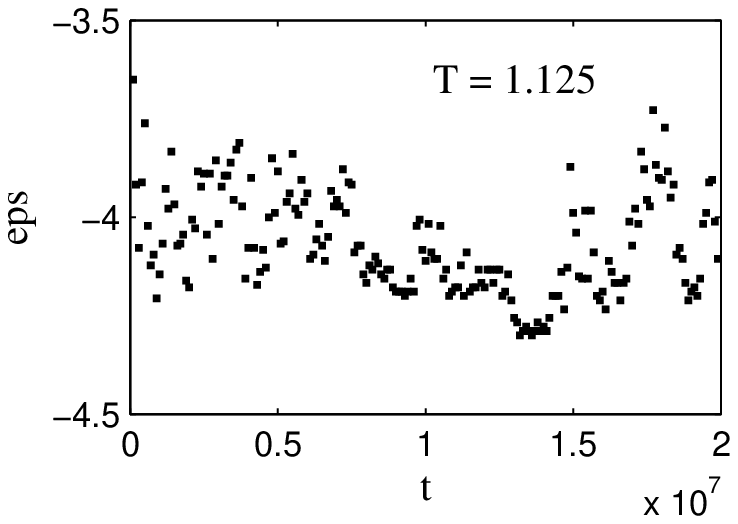}
\hspace{0.1cm}
\includegraphics{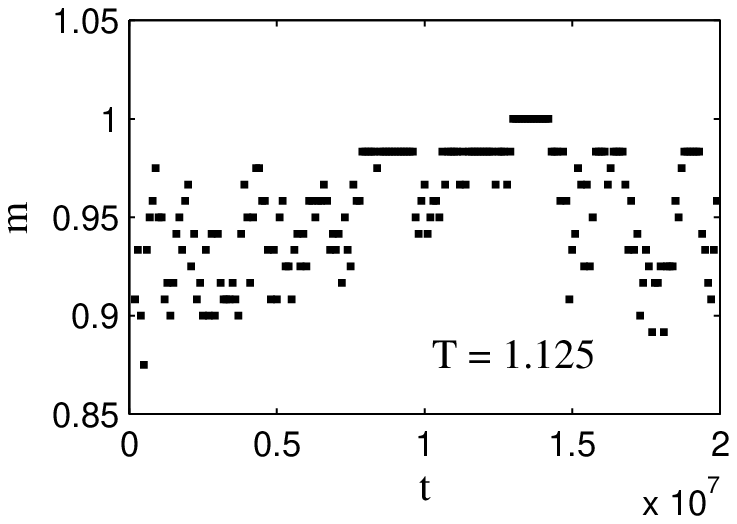}

\includegraphics{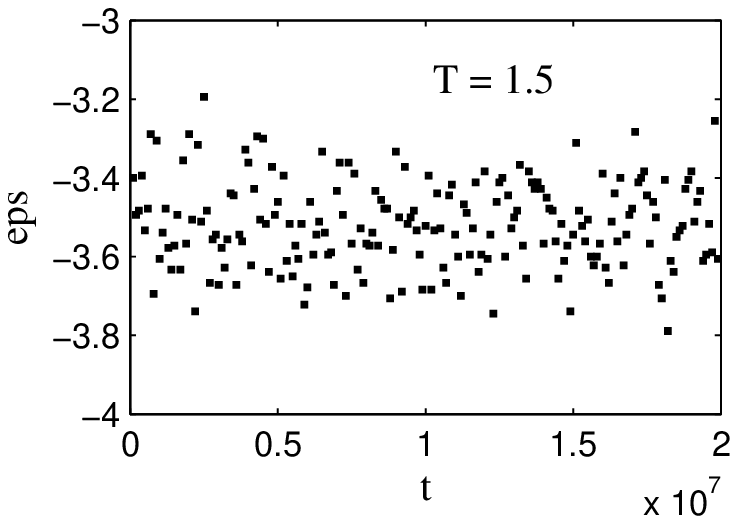}
\hspace{0.1cm}
\includegraphics{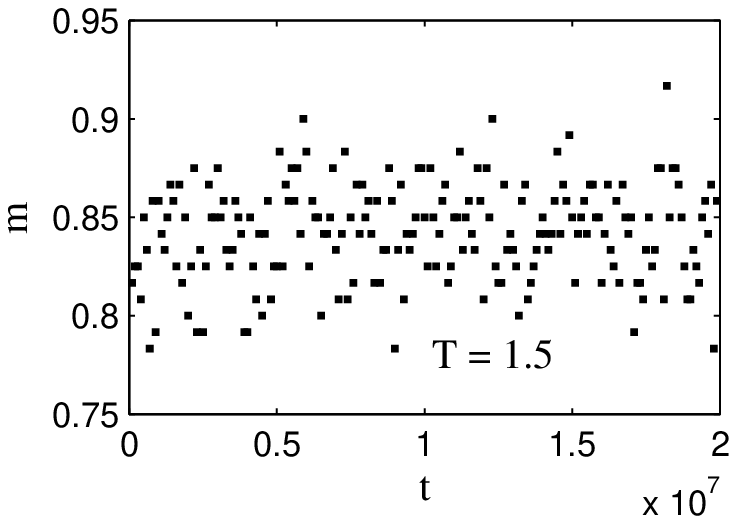}
\end{center}
\caption{Examples of plots that were used to estimate equilibration times $t_{\textrm{eq}}$:
energy per edge $\epsilon$ and $m$ versus simulation time $t$ for $N_e = 180$ edges.}
\label{relaxation}
\end{figure}

Figure \ref{relaxation} contains examples of plots used
to estimate the equilibration times for different temperatures $T$ and sizes
$N_e$ of the system.
The plots show data from preparatory simulations that measured the
energy per edge $\epsilon = E / N_e$ and the observable $m = M / N_e$
for a duration of $t = 2\cdot 10^7$ Monte Carlo steps.
The starting configuration is a graph of $N_e$ disconnected edges

It should be noted how the characteristics of the plots change as we
go from low to higher temperatures.
The plot is very autocorrelated for $T = 0.5$,
becomes less correlated and more fluctuating at $T = 1.125$,
and turns into strong fluctuations with little autocorrelation at $T = 1.5$.

Based on these plots and similar ones for other temperatures
and edge numbers, I adopted a set
of equilibration times $\Delta t_{\textrm{eq}}$ for the main simulation
which are listed in table \ref{tableequilibration}.
The total number of Monte Carlo steps $t_{\textrm{\tiny MC}}$ is taken
to be $100 \Delta t_{\textrm{eq}}$ in each case.

\psfrag{eps}{\large $\epsilon$}
\psfrag{m}{$m$}
\psfrag{t}{$t$}
\psfrag{T = 0.5}{$T = 0.5$}
\psfrag{T = 1.125}{$T = 1.125$}
\psfrag{T = 1.5}{$T = 1.5$}

\begin{table}
\begin{tabular}{l@{\quad}|@{\quad}l@{\quad}|@{\quad}r@{\quad\quad}r}
& $T$ & $\Delta t_{\textrm{eq}}$ & $t_{\textrm{\tiny MC}}$ \\ \hline
$N_e = 45$ & 0.5 -- 2.0 & $7\cdot 10^6$ & $7\cdot 10^8$ \\
& 2.25 -- 6.75 & $1\cdot 10^6$ & $1\cdot 10^8$ \\ \hline
$N_e = 90$ & 0.5 -- 2.0 & $7\cdot 10^6$ & $7\cdot 10^8$ \\
& 2.25 -- 6.75 & $1\cdot 10^6$ & $1\cdot 10^8$ \\ \hline
$N_e = 180$ & 0.5 -- 1.5 & $2\cdot 10^7$ & $2\cdot 10^9$ \\
& 1.625 -- 2.5 & $7\cdot 10^6$ & $7\cdot 10^8$ \\
& 2.75 -- 6.75 & $1\cdot 10^6$ & $1\cdot 10^8$
\end{tabular}
\caption{Equilibration times and total number of Monte Carlo steps.}
\label{tableequilibration}
\end{table}

\subsection{Evidence for a phase transition?}
\label{evidenceofaphasetransition}

% curves for \epsilon, different N's together
% curves for m, different N's together

\psfrag{eps}{\large $\epsilon$}
\psfrag{T}{$T$}
\psfrag{Ne = 45}{$N_e = 45$}
\psfrag{Ne = 90}{$N_e = 90$}
\psfrag{Ne = 180}{$N_e = 180$}
\begin{figure}
\begin{center}
\includegraphics{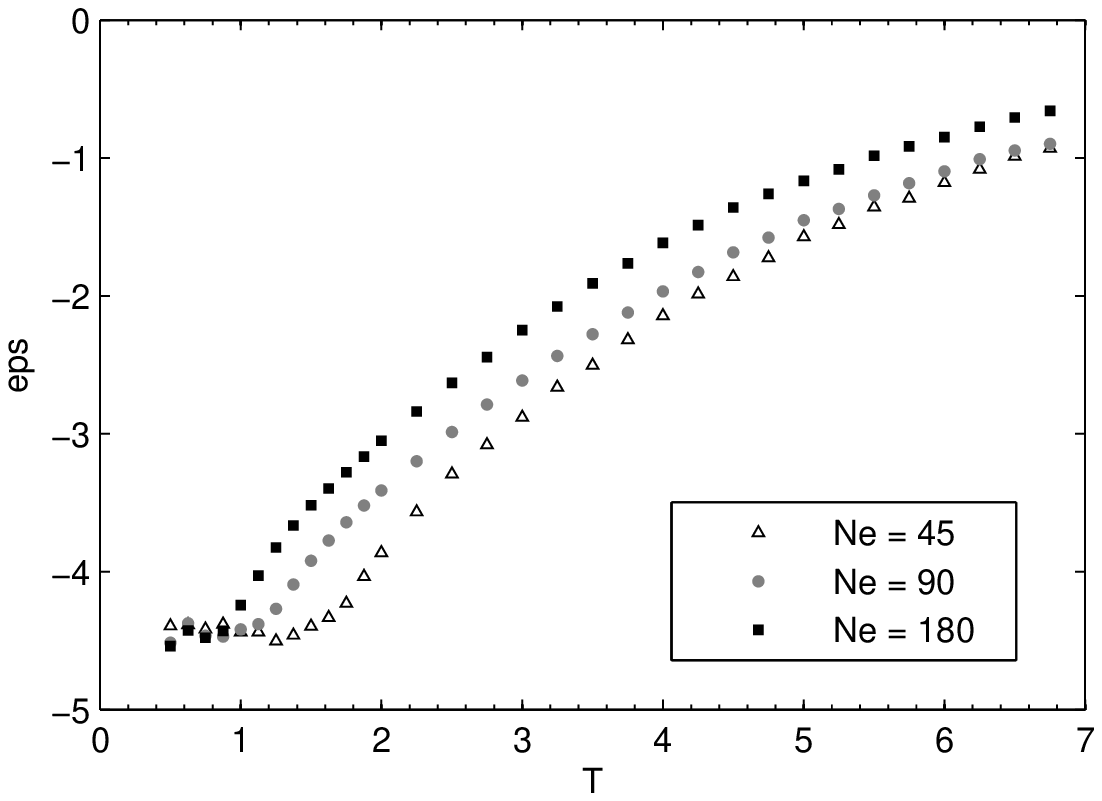}
\vspace{1.0cm}

\includegraphics{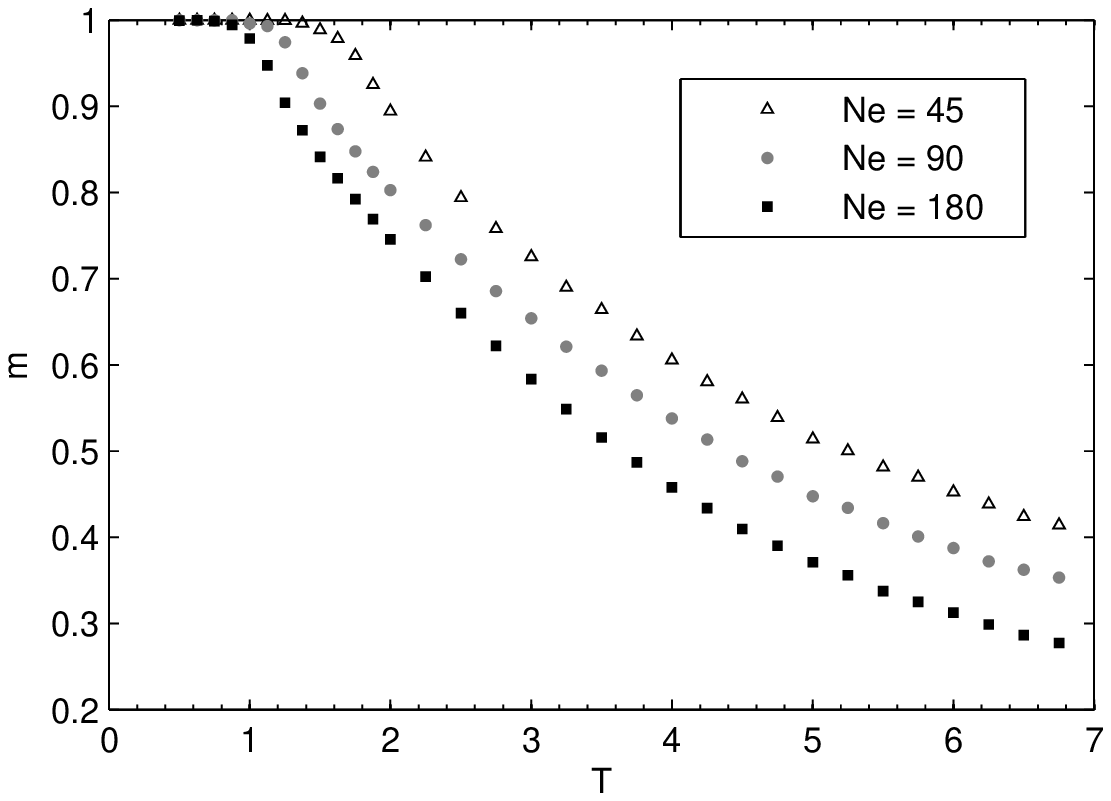}
\end{center}
\caption{Energy per edge $\epsilon$ and $m$ versus temperature $T$ for edge number $N_e = 45$, $90$ and $180$.}
\label{epsilonandm}
\end{figure}

Figure \ref{epsilonandm} displays the simulation results for the energy per edge $\epsilon$ and the observable $m$.
Here and in the following error bars are omitted whenever they are smaller than the symbols of the data points\footnote{Correlation
times and error estimates will be addressed in section \ref{subtletiesatlowtemperature}.}
At sufficiently low temperature $m$ reaches the value 1 and, up to conical singularities, the average graph
is given by a surface without boundary.
This confirms the expectations expressed in the previous section.
The attainment of $m = 1$ goes along with the energy per edge reaching its bottom value.
(Curiously, for $N_e = 45$, the energy rises again slightly as $T$ decreases further.
This might be related to incomplete equilibration at low temperatures, see subsec.\ \ref{subtletiesatlowtemperature}).
The transition temperature, where $m = 1$ sets in, decreases as $N_e$ increases.

% curves for c, including derivative of fit to E
% separately for different N's

\psfrag{c}{\large $c$}
\psfrag{from var 123}{$\beta^2 N_e \left(\b \epsilon^2\ket - \b \epsilon\ket^2\right)$}
\psfrag{from deriv}{$\d \b \epsilon\ket / dT$ (from fit)}

\begin{figure}
\vspace{-1cm}

\begin{center}
\includegraphics{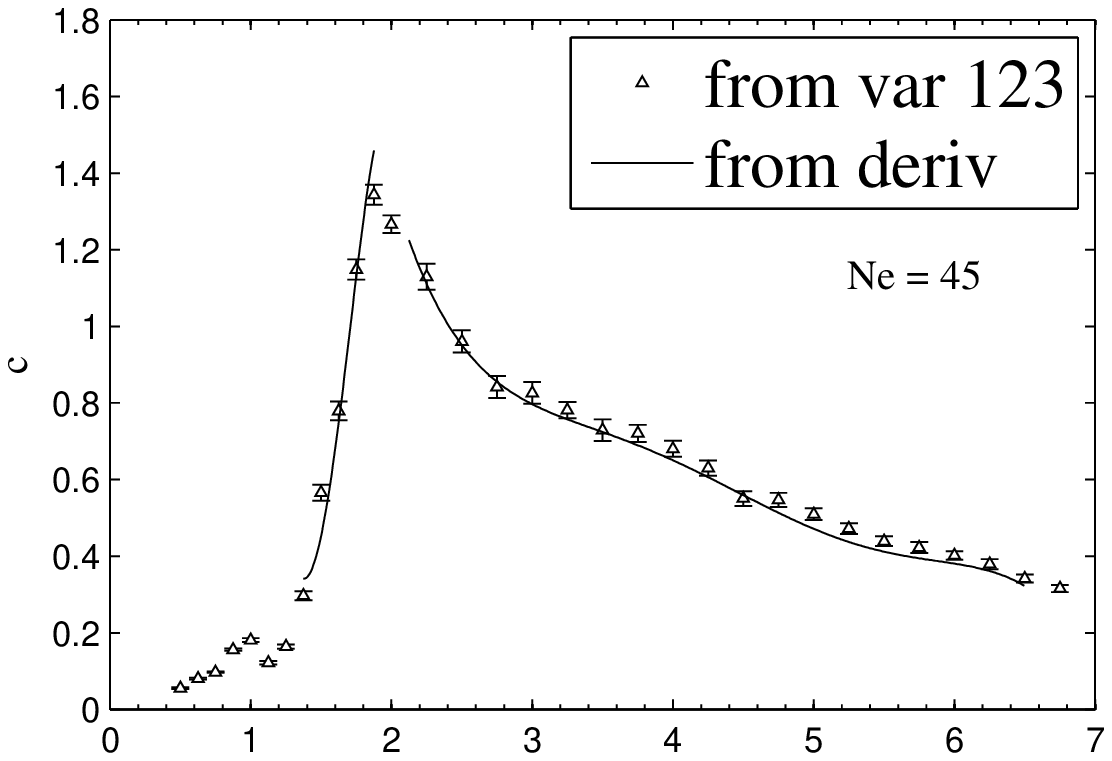}
%\vspace{1.0cm}

\includegraphics{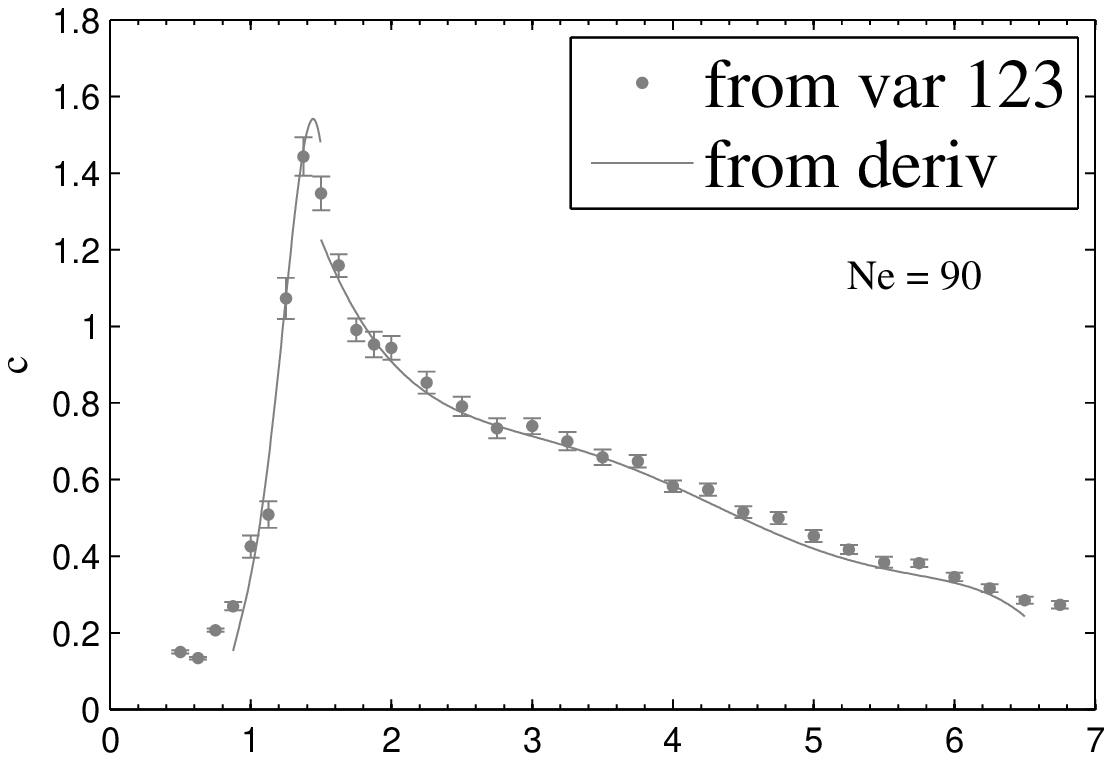}
%\vspace{1.0cm}

\includegraphics{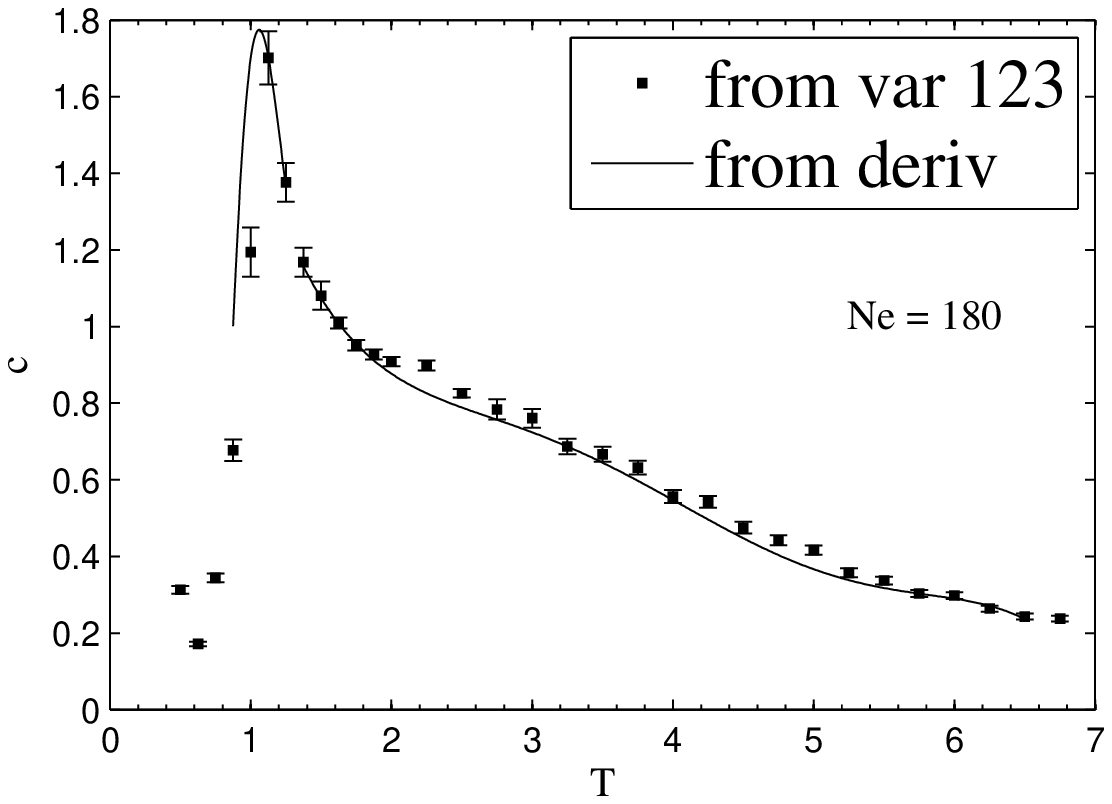}
\end{center}
\caption{Specific heat per edge $c$ versus temperature $T$ for $N_e = 45$, $90$ and $180$.}
\label{specificheat}
\end{figure}

That something interesting happens when $m$ becomes 1
is also suggested by the results for the specific heat,
shown in \fig{specificheat}.
The specific heat can, in principle, be determined in two ways;
from the variance,
\be
c = \beta^2 N_e \left(\b \epsilon^2\ket - \b \epsilon\ket^2\right)\,,
\label{cfromvariance}
\ee
or from the temperature derivative of the average energy per edge $\b \epsilon\ket$:
\be
c = \d \b \epsilon\ket / dT
\label{cfromderivative}
\ee
In theory these formulas should give the same value,
and the same should hold for measurements in the simulation,
if the sampling is representative of the equilibrium.

In \fig{specificheat} the symbols represent the values of $c$ obtained
via formula \eq{cfromvariance}.
The errors are estimated by the blocking method.
The continuous lines, on the other hand, come from
fitting polynomials to the $\epsilon$--$T$ plots in \fig{epsilonandm}
and then taking the deriviative with respect to $T$.
We see that in the temperature range of the fits
there is relatively good agreement between the two methods,
thus suggesting that the simulation is equilibrated.

\begin{figure}
\begin{center}
\includegraphics{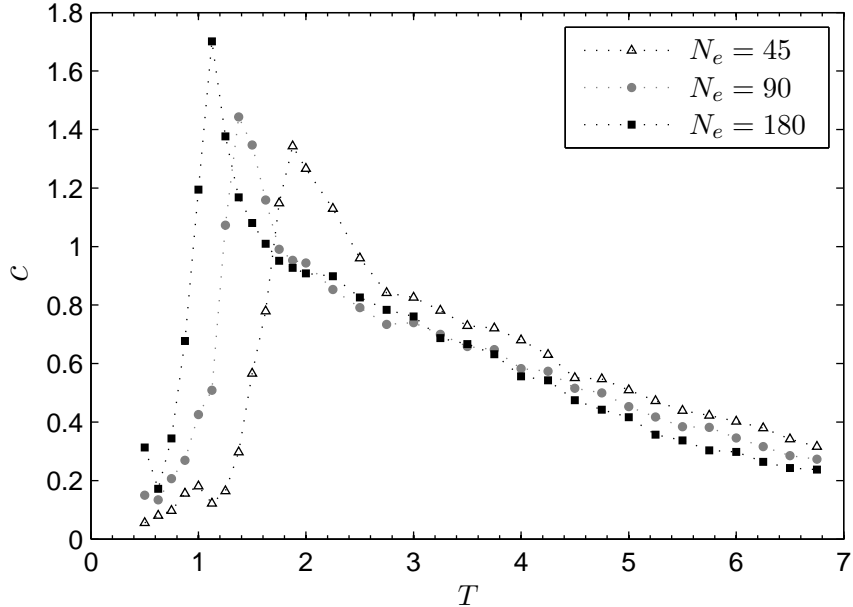}
\end{center}
\caption{Specific heat per edge $c$ versus temperature $T$.}
\label{specificheatallN}
\end{figure}

The most notable feature of the plots are the peaks which reside
near the temperature where $m$ reaches 1.
For $N_e = 45$, $90$ and $180$ the peaks lie at
$T = 1.875$, $1.375$ and $1.125$ respectively\footnote{It is
not an accident that these values are close to 1.
When choosing the couplings in the Hamiltonian, I tuned them to the onset of surface formation,
while $T$ was set to 1.}.
For easier comparison, the curves for different $N_e$
are plotted together in \fig{specificheatallN}.
The dotted line was added for clarity and connects the points
by straight lines.

% curves for c, all N's together

\begin{figure}
\begin{center}
\includegraphics{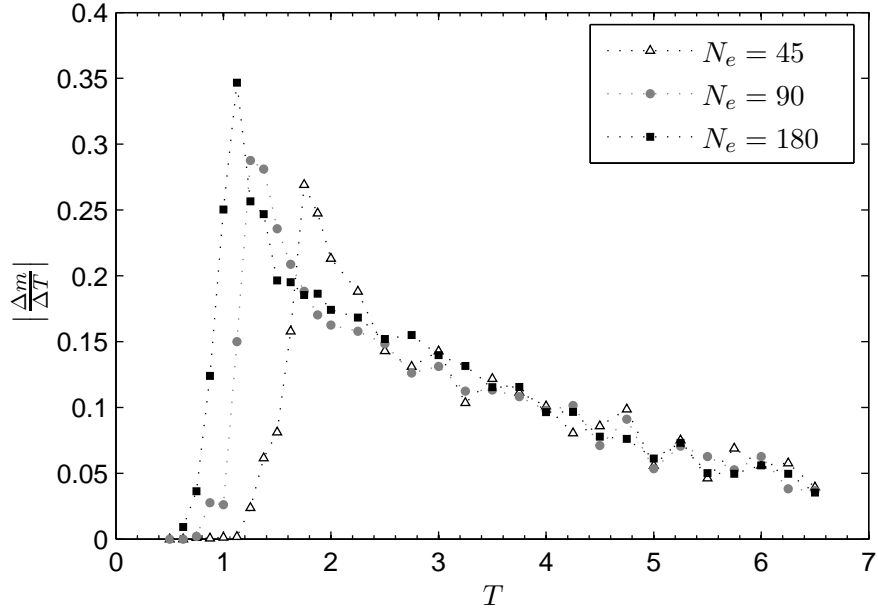}
\end{center}
\caption{Forward derivative of measured values of $m$ with respect to temperature $T$.}
\label{discretederivativem}
\end{figure}

This plot should be compared with \fig{discretederivativem},
which was obtained by taking the discrete forward derivative
of the points in the $m$--$T$ plot.
The peaks of $\left|\frac{\Delta m}{\Delta T}\right|$ are slightly
shifted relative to those of $c$. This may be an artifact resulting
from the use of the forward derivative.

% derivative m, all N's together

\psfrag{deriv m}{$\left|\frac{\Delta m}{\Delta T}\right|$}

\begin{figure}
\begin{center}
\includegraphics{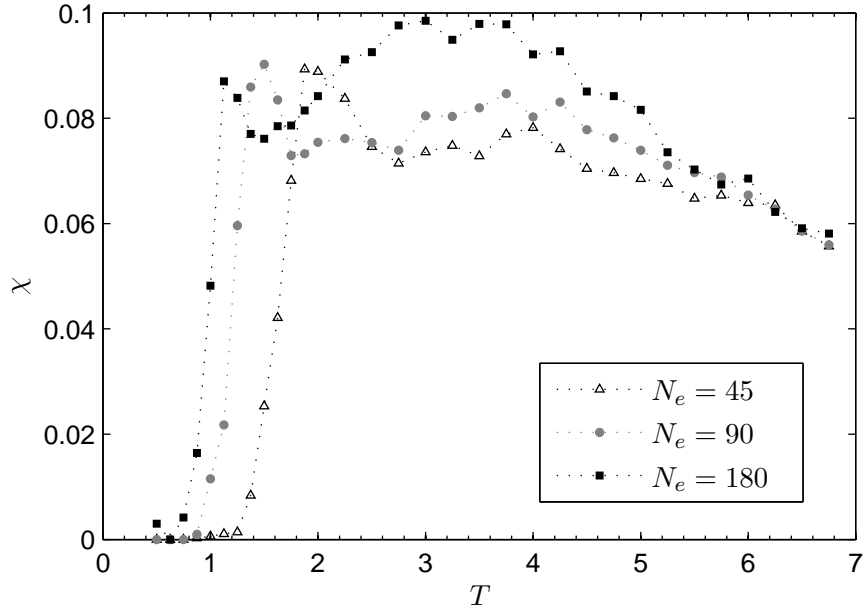}
\end{center}
\caption{Susceptibility $\chi$ versus temperature $T$.}
\label{susceptibility}
\end{figure}

Figure \ref{susceptibility} shows the results for the susceptibility $\chi$
defined in eq.\ \eq{definitionsuseceptibility}.
$\chi$ exhibits a sharp increase at the transition temperature.
Unlike for a ferromagnet, there is no peak, however, since
$\chi$ stays more or less constant, as $T$ is further increased.

% curves for \chi, all N's together

\psfrag{chi}{$\chi$}

The salient properties of our data can be summarized as follows.
Below a transition temperature $T_{\textrm{trans}}$ the surface observable $m$ is constant
and equals 1. At the temperature$T_{\textrm{trans}}$, $m$ starts to decrease.
The rate of this decrease grows as $N_e$ increases. That is,
the kink in the $m$--$T$ curve becomes more pronounced as the system
becomes larger.
At the transition temperature $T_{\textrm{trans}}$, we also see a peak in the specific heat $c$
whose amplitude increases with system size.
The peak in $c$ is accompanied by a notable increase in the time
$\Delta t_{\textrm{eq}}$ needed to equilibrate the system.
A further indicator is the ``susceptibility'' $\chi$ which shows a sharp rise
at $T = T_{\textrm{trans}}$.

One can infer from this that for $T < T_{\textrm{trans}}$ the system
is characterized by surface--like graphs for which $m = 1$.
For $T > T_{\textrm{trans}}$, these surfaces are replaced by more general graphs
with $m < 1$. The observable $m$ serves as a parameter to distinguish
these two regimes.

The $T$--dependence of the specific heat is similar to what is found
in more conventional systems with phase transitions.
It should be noted, however, that the peak of the specific heat shifts to
lower temperatures, as the system size increases.
Thus, it is possible that the transition temperature $T_{\textrm{trans}}$ goes to zero
in the infinite size limit.
Given that I have measured three sizes up to $N_e = 180$ so far,
it is premature to conclude whether this is the case or not.
That is, one cannot tell at this stage if the system exhibits a phase transition
at zero $T$ or at finite $T$.
Further remarks on this will be made in the conclusion of the paper.

\subsection{Surface properties}
\label{surfaceproperties}

This subsection deals in more detail with the surface states found at low temperature.
What kind of surfaces appear? What defects do they have? And how do they change when
$T$ is increased?

\begin{figure}
\begin{center}
\includegraphics{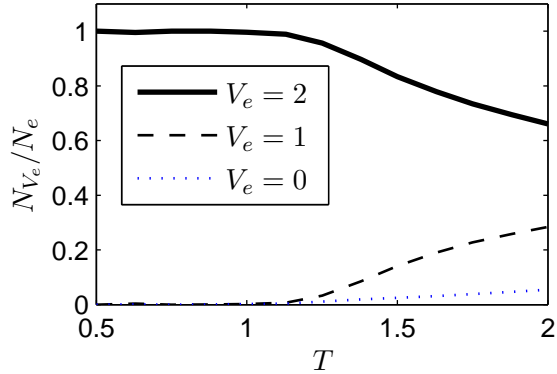}
\end{center}
\caption{The fraction of edges with valence $V_e$ versus the temperature $T$ for $N_e = 90$.}
\label{valencofedges}
\end{figure}

To answer these questions I conducted two additional sets of simulations.
The first kind of simulation was aimed at showing more precisely how the valence of edges evolves
as the temperature is raised above the transition temperature.
This led to the plots in \fig{valencofedges}. For $T < T_{\textrm{trans}}$ almost all edges have valence $V_e = 2$.
At $T = T_{\textrm{trans}}$ the fraction of these edges starts to decrease until it drops to around 66 percent at $T = 2.0$.
In parallel the fraction of boundary edges ($V_e = 1$) rises from 0 percent to about 28 percent at $T = 2.0$.
The percentage of edges outside of triangles ($V_e = 0$) increases as well and reaches 5 percent.
It is clear from this that at $T = 2.0$ the graphs must have lost most of its resemblance with a surface,
since a considerable fraction of edges form boundaries of triangles.

\psfrag{NVe / Ne}{$N_{V_e} / N_e$}
\psfrag{Ve = 0}{$V_e = 0$}
\psfrag{Ve = 1}{$V_e = 1$}
\psfrag{Ve = 2}{$V_e = 2$}

The second type of simulation concerned defects, topology and curvature.
The measurements proceeded in three steps. First, the conical singularities and the
edges with zero valence are determined and counted. Then, these defects are removed,
so that one obtains a reduced graph that is equivalent to a triangulation of a surface.
This reduction to a surface is needed in order to have a well--defined notion of topology.
In the third step, we measure various quantities related to topology, boundary size and
curvature of the surface.

The curvature is obtained by interpreting the triangulation as a piecewise--linear manifold whose
edges have length 1. The deficit angle at a vertex is given by
\be
\delta_v = 2\pi - V_v \pi / 3
\ee
and the Ricci scalar is defined by
\be
R_v = \frac{2\delta_v}{A_v}\,.
\ee
Here, $A_v = V_v A / 3$ is the area associated to the vertex, and $A = \sqrt{3} / 4$ is the area of the triangles
of the triangulation (see e.g.\ \cite{JankeJohnstonWeigel}).

\begin{table}
\begin{minipage}{8.3cm}
\begin{tabular}{r@{\quad}|@{\quad}l@{\quad}|@{\quad}l}
$T$ & 0.625 & 2.0 \\ \hline
$N_{V_e = 0} / N_e$ & 0.00 $\pm$ 0.00 & 0.09 $\pm$ 0.01 \\
$N_{\textrm{con}} / N_v$ & 0.08 $\pm$ 0.01 & 0.40 $\pm$ 0.02 \\ \hline
$N_{cc}$ & 1.2 $\pm$  0.1 & 2.6 $\pm$ 0.3 \\
$V_v$ & 6.7 $\pm$ 0.1 & 5.3 $\pm$ 0.1 \\
$R$ & -0.8 $\pm$ 0.1 & 3.0 $\pm$ 0.3 \\ \hline
$N^{lc}_{be} / N^{lc}_e$ & 0.00 $\pm$ 0.00 & 0.36 $\pm$ 0.01 \\
$\textrm{orientability}$ & 0.00 $\pm$ 0.00 & 0.07 $\pm$ 0.01 \\
$\chi_{\ssst\textrm{Euler}}$ & -6.1 $\pm$ 0.4 & -4.8 $\pm$ 0.4
\end{tabular}
\end{minipage}
\hspace{0.5cm}
\begin{minipage}{7cm}
\begin{tabular}{l@{\quad}l}
$N_v$ & \# of vertices \\
$N_e$ & \# of edges \\
$N_{V_e = 0}$ & \# of edges with valence $V_e = 0$ \\
$N_{\textrm{con}}$ & \# of conical singularities \\
$N_{cc}$ & \parbox[t]{5cm}{\raggedright \# of connected components \\ of surface}  \\
$V_v$ & (average) valence of vertices \\
$R$ & (average) Ricci scalar \\
$N^{lc}_e$ & \# of edges in largest component \\
$N^{lc}_{be}$ & \parbox[t]{5cm}{\raggedright \# of boundary edges in \\ largest component}
\end{tabular}
\end{minipage}
\caption{Surface properties for edge number $N_e = 180$ and temperatures $T = 0.625$ and $T = 2.0$.
The table consists of three sections which refer to the graph, the surface (defined by the reduced graph)
and the largest component in the surface.}
\label{tabletopology}
\end{table}

Since these measurements involve a change in the graph (and saving graphs was not implemented in the code),
I ran the simulation differently from the previous ones.
Instead of a sequence start $\rightarrow$ equilibrate $\rightarrow$ measure $\rightarrow$ wait $\rightarrow$ measure \ldots,
I used a schedule that restarts the simulation after each measurement:
\begin{quote}
start $\rightarrow$ anneal $\rightarrow$ equilibrate $\rightarrow$ measure \\
$\rightarrow$ restart $\rightarrow$ anneal $\rightarrow$ equilibrate $\rightarrow$ measure \ldots
\end{quote}
A similar simulation schedule will also play a role in the next section, where I discuss equilibration at
low temperature.

The results are given in table \ref{tabletopology}.
I ran the simulation for $T = 0.625$ and $T = 2.0$ to compare the graph properties
below and above the critical temperature.
At low temperature, the graph contains no edges with valence $V_e = 0$.
Eight percent of the vertices come with conical singularities.
The typical surface consists of a single connected component
with moderate curvature.
The largest component of the surface has no boundary,
it is non--orientable and on average given by a connected
sum of 8 real projective spaces. (Recall that a non--orientable surface with
Euler characteristic $\chi_{\ssst\textrm{Euler}} = 2 - k$ is a connected sum of $k$ real projective planes.)

At the temerature $T = 2.0$ above the transition point,
the number of edges outside triangles is 9 times larger
than in the surface regime, and the fraction of conical
singularities reaches 40 percent.
While we can still define a reduced graph and hence a surface,
it is no longer a surface in the usual sense of the word.
In the largest component, 30 to 40 percent of the edges are
boundary edges.

Let me add that I also tried to measure the spectral dimension of the graphs.
This attempt failed, however.
When measuring the return probability of diffusion,
there is typically a window in the diffusion time that can be
used to determine the spectral dimension. For diffusion times outside
this interval the measurement is hampered by discretization
and finite volume effects.
In the present case, the system size appears to be so small ($N_e \le 180$)
that discrete and finite volume effects overlap, and no clear signal
of the spectral dimension is visible.

\section{Subtleties at low temperature}
\label{subtletiesatlowtemperature}

The simulation of systems with complex energy structure and many degeneracies
can pose considerable challenges, especially at low temperatures.
Well--known examples for this are systems with quenched disorder
such as spin--glasses and random field systems
(see e.g.\ \cite{Young,HarmannRicciTersenghi,HartmannFindingLowtemperaturestates}).
The simulation may get trapped in certain configurations,
and never find the bottleneck to exit and enter other regions.
It may sample some of the ground states, but leave out others,
and thus create biased results that are not representative of the true
equilibrium.

% plot, acceptance rate for T = 0.5 and T = 2.0
\psfrag{T = 2.0}{$T = 2.0$}
\psfrag{R}{$R_a$}
\begin{figure}
\begin{center}
\includegraphics{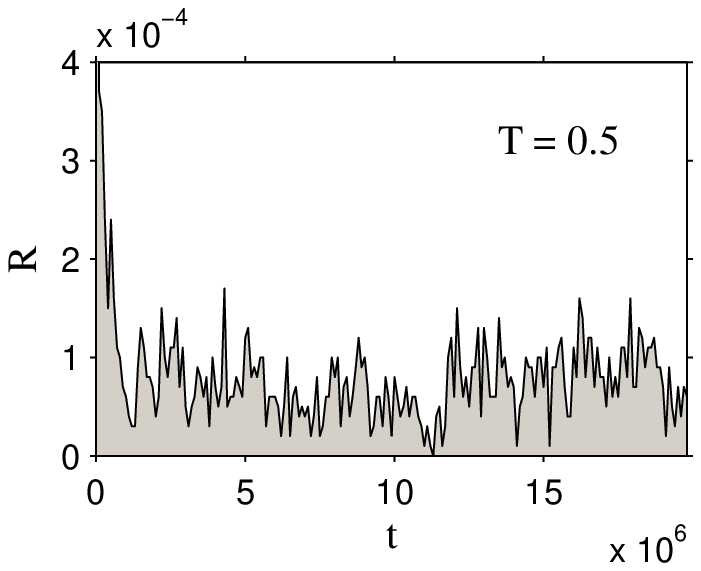}
\hspace{0.02cm}
\includegraphics{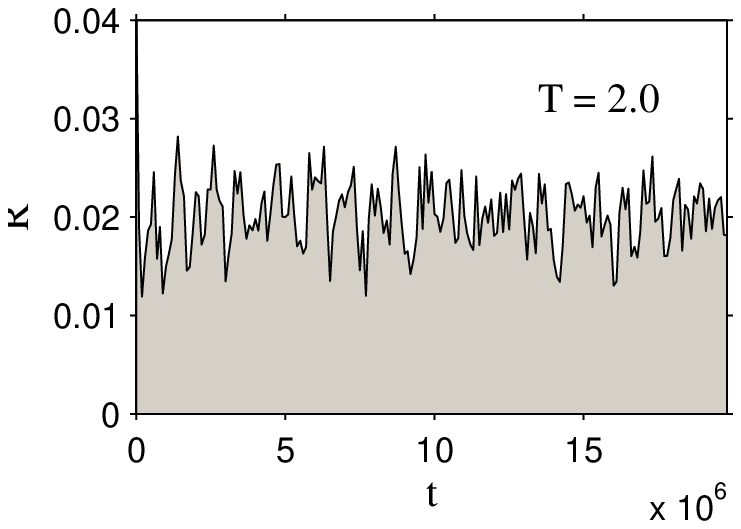}
\end{center}
\caption{Acceptance rate $R_a$ versus step number $t$ for $T = 0.5$ and $T = 2.0$.}
\label{acceptancerates}
\end{figure}

The ground states of the present model form a very special subset of the entire configuration space.
Therefore, it is not far--fetched to imagine that similar difficulties could occur here.
A first indication of this problem can be seen in the very small acceptance rates
at low temperature. While the acceptance is already low, in general, it drops to values
below $10^{-3}$ as we reach $T = 0.5$ (see \fig{acceptancerates}).
Real--time visualization with the Ubigraph software revealed that at this temperature
the accepted changes are predominantly 2--2 Pachner moves---moves that are not able
to change the topology of a triangulation.

\begin{figure}
\begin{center}
\parbox{2cm}{
\hspace{-2.5cm}\includegraphics{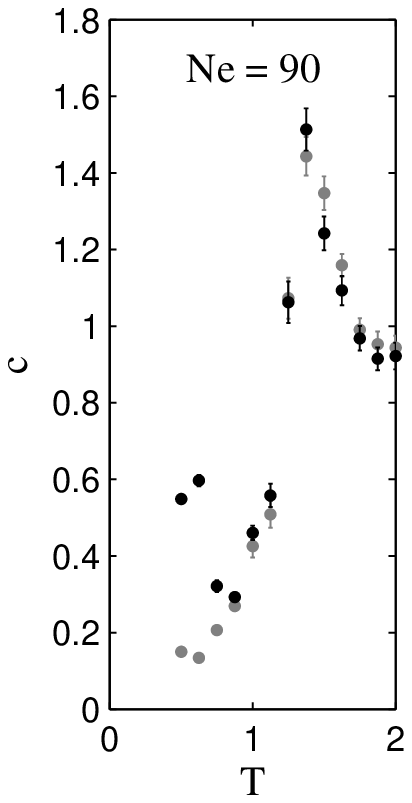}
}
\parbox{10cm}{
\parbox{11.8cm}{
\parbox{7.6cm}{
\includegraphics{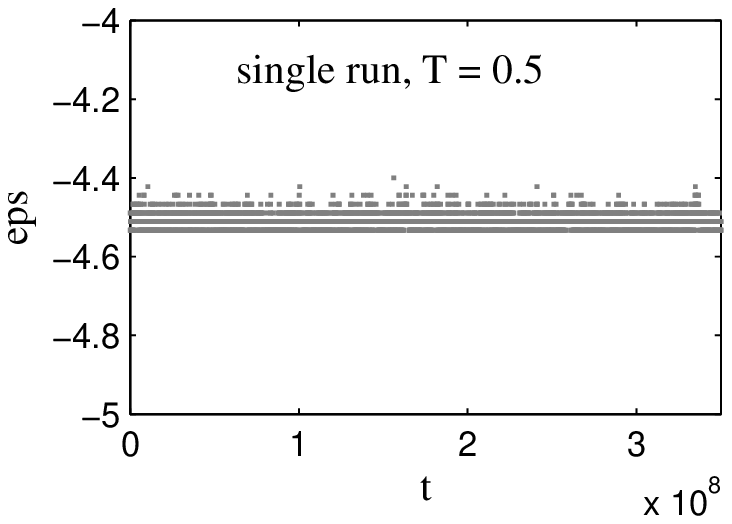}
}
\fbox{
\parbox{3cm}{
\begin{tabular}{l@{\quad}l}
$\Delta t_{\textrm{\tiny eq}}$ & $7\cdot 10^6$ \\
$\Delta t_{\textrm{\tiny measure}}$ & $7\cdot 10^4$ \\
$t_{\textrm{\tiny MC}}$ & $7\cdot 10^8$
\end{tabular}
}
}
}
\parbox{11.8cm}{
\parbox{7.6cm}{
\includegraphics{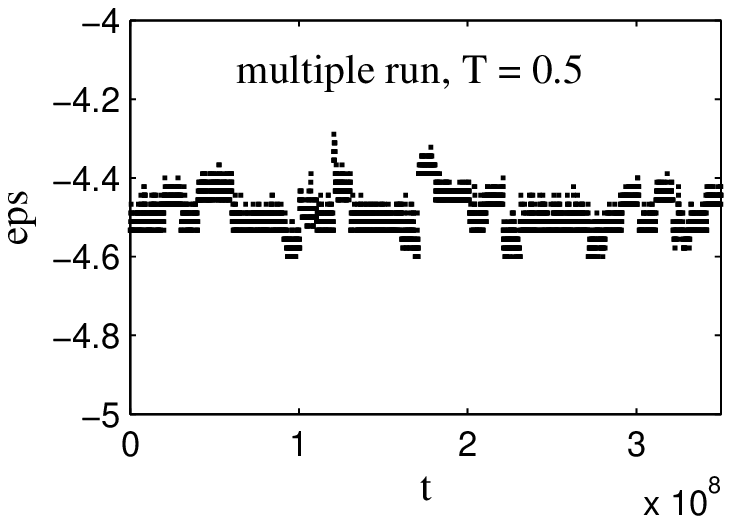}
}
\fbox{
\parbox{3cm}{
\begin{tabular}{l@{\quad}l}
$\Delta t_{\textrm{\tiny anneal}}$ & $3\cdot 10^6$ \\
$\Delta t_{\textrm{\tiny eq}}$ & $7\cdot 10^6$ \\
$\Delta t_{\textrm{\tiny measure}}$ & $5\cdot 10^4$ \\
$t_{\textrm{\tiny MC}}$ & $20\cdot 10^6$ \\
$N_{\textrm{\tiny repeat}}$ & 35
\end{tabular}
}
}
}
}
\end{center}
\caption{Comparison of the standard and alternate simulation schedule:
on the left, specific heat $c$ versus temperature $T$ for
standard schedule (grey) and modified schedule (black);
in the middle, plot of $\epsilon$ vs.\ simulation time $t$;
on the right, parameters of Monte Carlo simulation.}
\label{comparisonofschedules}
\end{figure}

The question is therefore the following.
Can the simulation sample the low--energy states in a statistically correct way,
or does it get stuck in a minimum energy state and stay there for the rest of the
simulation?
To obtain more information about this, I repeated some of the previous simulations with a different
schedule. In this new schedule, a phase of annealing and equilibration is followed by a sequence of measurements,
then the simulation restarts and the whole process is repeated.
This procedure may yield a better sampling of
ground states and shorten the time during which the algorithm is trapped.
The precise parameters of the original and modified schedule are indicated
in \fig{comparisonofschedules}. During annealing the temperature starts at
$T = 4.0$ and drops gradually to the target temperature.

% plot, comparing T = 0.5 to 2.0 with single run and mixed run
\psfrag{single run, T = 0.5}{standard schedule, $T = 0.5$}
\psfrag{multiple run, T = 0.5}{alternate schedule, $T = 0.5$}

The two types of simulations agree reasonably well for $T \ge 1.0$. For the lowest temperatures, however,
the alternate schedule leads to a much higher value of the specific heat.
The reason for this difference becomes clear when we inspect the measured data for the energy per edge (see \fig{comparisonofschedules}).
In the standard case, the fluctuations exhibit a fixed pattern for the length of the entire simulation.
In the alternate case, the same kind of pattern appears, however, only for the duration of a single
measurement period. At the beginning of the following measurement sequence, the pattern jumps to a new energy
and remains there until the next restart.

% autocorrelation plots
\psfrag{t}{$t$}
\psfrag{a(t)}{$a(t)$}
\psfrag{T = 0.5}{\small $T = 0.5$}
\psfrag{T = 0.75}{\small $T = 0.75$}
\psfrag{T = 1.0}{\small $T = 1.0$}
\psfrag{T = 1.25}{\small $T = 1.25$}
\psfrag{T = 1.125}{$T = 1.125$}
\begin{figure}
\begin{center}
\includegraphics{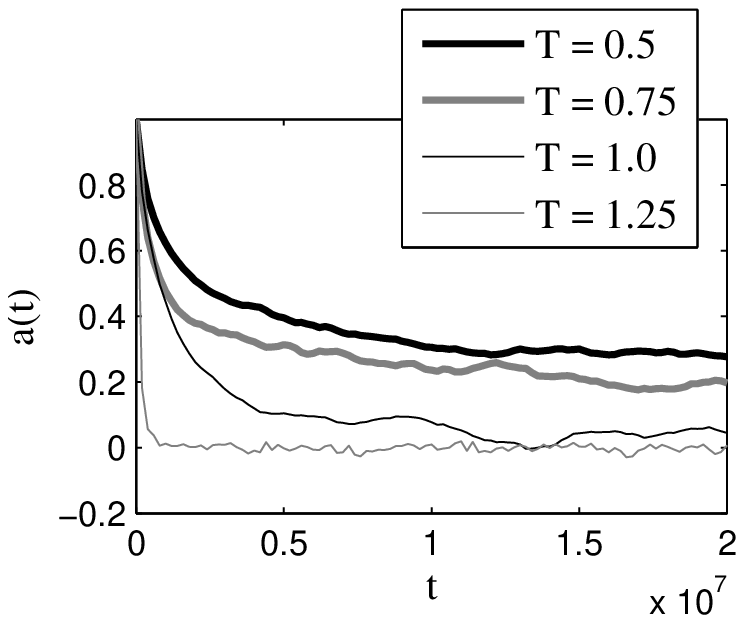}
\hspace{0.1cm}
\includegraphics{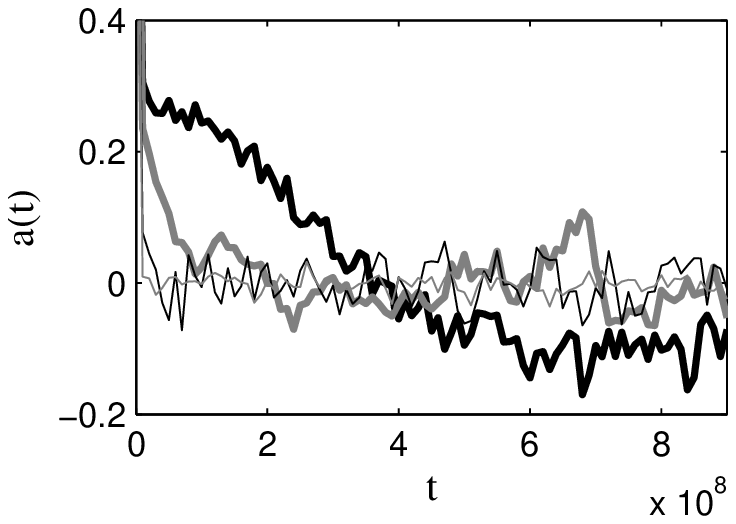}

\includegraphics{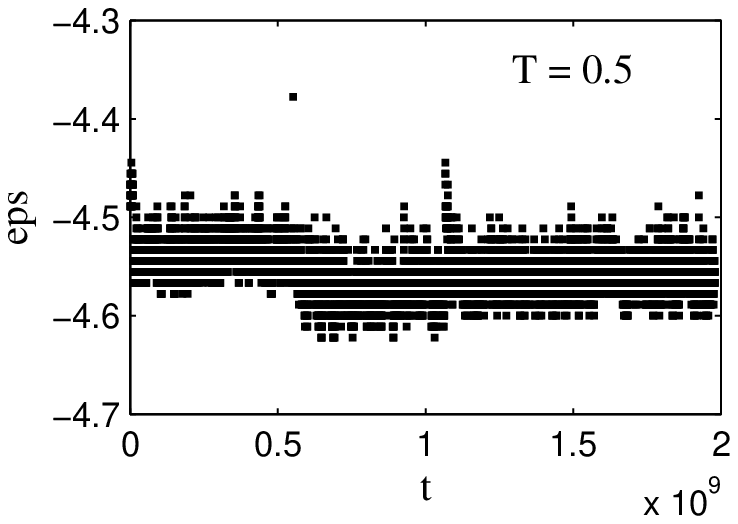}
\hspace{0.1cm}
\includegraphics{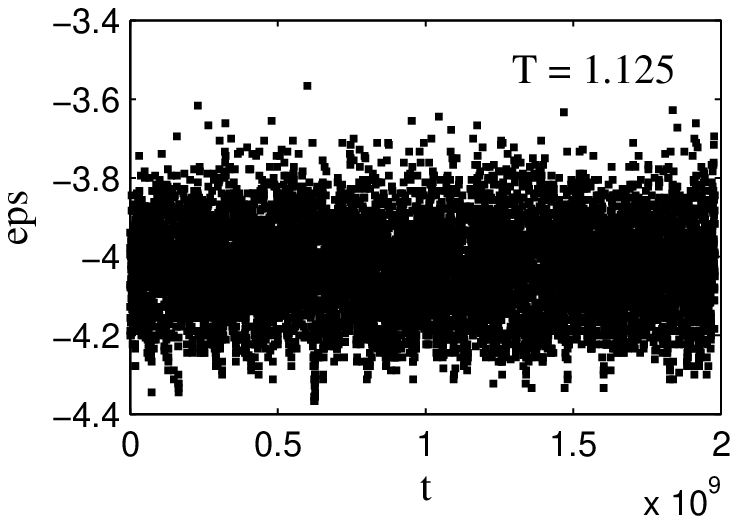}
\end{center}
\caption{Autocorrelation function $a(t)$ of $\epsilon$ for different temperatures and $N_e = 180$.}
\label{autocorrelation}
\end{figure}

This suggests the following interpretation.
The standard schedule gets stuck in one set of states and stays there for the rest of the simulation.
The alternate schedule gets trapped as well, but only for a short period and each time for a different energy.
This sampling of different states results in a too large variance and hence in a wrong value for the
specific heat.
Within this sample, however, there are also states that have lower energy than the states found by the
conventional scheme. Thus, there is something wrong and right in both methods.
The original simulation yields the correct specific heat, but the measured energy may be too high,
if the algorithm freezes in a set of states above the true ground state.

The last point to be addressed is the autocorrelation of our data (see \fig{autocorrelation}).
For temperatures greater than $T_{\textrm{trans}}$, the autocorrelation drops exponentially to zero and I
determined corresponding correlation times $\Delta t_{\textrm{cor}}$ for the error analysis. For $T < T_{\textrm{trans}}$,
on the other hand, the data are, in many cases, autocorrelated on time scales of the order of $t_{\textrm{\tiny MC}}$,
so I would not be able to correct this other than by going to much larger simulation times.
The plots show furthermore that my estimates for the equilibration time were too optimistic at low temperatures.

\section{Conclusion}

Let me summarize the results.
I have defined a graph model that produces ``space from no space''.
Two--dimensional triangulations (space) arise as ground states from a much
larger set of configurations (no space).
To measure the extent to which a graph is a surface I introduced an observable
called $m$.
Monte Carlo simulations show that below a transition temperature
$T_{\textrm{trans}}$ the observable $m$ is close or equal to 1,
and the graphs form triangulations of surfaces\footnote{modulo conical singularities}.
For temperatures $T > T_{\textrm{trans}}$, one has $m < 1$ and the surfaces are
replaced by more general graphs.
More detailed measurements reveal that the surfaces are typically
connected and non--orientable.
As we go above the transition temperature,
the graphs lose the surface structure, since the number of defects increases
and many of the edges become boundaries of triangles.

The temperature dependence of $m$ and of the specific heat $c$
suggest that there is a phase transition between the regime
with and without surfaces.
However, the transition temperature $T_{\textrm{trans}}$ drifts, as the
system size is increased, so it is possible that $T_{\textrm{trans}}$
goes to zero in the infinite--size limit.
In this case, there would be no phase transition at finite temperature,
and one would instead speak of a \textit{zero temperature phase transition}.
Since the system was only simulated for very few and small sizes, it is
too early to tell which scenario is realized.

What are future directions for this work?
Obviously, one should try to extend the model to larger
sizes and higher, more realistic dimensions.
The key technical question in this will be whether one is able to
make the algorithm efficient enough to sustain equilibrium at low temperature.
It may be necessary to invent more sophisticated simulation schedules
in order to deal with larger edge numbers and higher--dimensional simplices.
Another remedy could be to use a different Hamiltonian that is more ``flexible''
in its choice of ground states. By this I mean that ground states
would not only be given by graphs that are literally 2d triangulations, but also by more
general graphs that are effectively 2--dimensional (e.g.\ in the spectral sense).
Perhaps one could define a ``spectral Hamiltonian'' that achieves this.

Furthermore, one should determine conclusively whether there is a phase transition,
and if it is at finite or zero temperature (in the infinite--size limit).
A phase transition at finite temperature is desirable in the sense
that it would make the behaviour near the transition universal
and reduce the amount of fine--tuning needed to achieve this state.
In the case of a zero temperature phase transition,
the dimensionless couplings would have to be sent to infinity as the system size
goes to infinity in order to maintain a transition at finite $T$.

Since the paper focused on the technical aspects, I would like to conclude
with a number of conceptual remarks.
An obvious question pertains to the role of time.
If this is a statistical model of space, there must be a time with respect
to which the system equilibrates.
If that is the case, it would seem that, in the model's world, time is of a different
nature than space, since space can dissolve, while there is always time for the
system to thermalize.

Let me give a possible interpretation of the model that clarifies this point.
Imagine that we had a consistent quantum theory of gravity that describes spacetime
as being built from elementary cells or building blocks.
Assume also that from a canonical point of view this spacetime
corresponds to blocks or ``atoms'' of space that evolve in a discrete, fluctuating time.
If the theory is a viable candidate theory of quantum gravity, there should be a regime where
classical continuum spacetime is recovered from the collective properties
of these atoms.
Like in conventional quantum theories, one might furthermore expect that there is
a statistical approximation of the theory, where the 4d path integral is
replaced by a statistical sum over 3d configurations.
Our model could be regarded as such a statistical approximation to a quantum
theory of spacetime (in this case 3d spacetime).
The edges of the graphs are the atoms of space and their interactions
are encoded by the Hamiltonian\footnote{Clearly, this is just a heuristic
consideration and leaves out many questions. What effect would fluctuations
of time have? How can Lorentz invariance be maintained?}.

This line of thought suggests also a connection with group field theory and
loop quantum gravity.
The quanta of these theories are triangles and tetrahedra and interact
to form pseudo--manifolds \cite{ReisenbergerRovelliFeynman1,ReisenbergerRovelliFeynman2,Oritigroupfieldtheoryapproach}.
If we were able to develop a statistical theory of group field theory and
loop quantum gravity, the resulting models may have a similar structure as
the model defined here\footnote{See \cite{RovelliVidotto} for another possible link between loop quantum gravity and statistical physics.}.

Apart from this, our system could be also useful as a concrete and simple
toy model that helps to imagine what a physics beyond spacetime could be;
a physics, that is, in which spacetime is no longer absolute
and instead only a special state among other states of the system.
This dethroning of spacetime is, in my view, a logical continuation of the development of
theoretical physics.
It is also a radical step that challenges accustomed ways of thinking about physics.
Let me highlight this by describing three thought experiments
that could be simulated with our model.

There is a 1d version of the Hamiltonian that results in the formation of 1d chains
at low temperature.
Imagine that we combine the 1d Hamiltonian with the 2d Hamiltonian used in this paper.
Then, one could vary the relative strength of the 1d and 2d couplings and
and observe transitions between a 1--dimensional and 2--dimensional
phase of the system.

As a second example, consider the surface regime of the model and suppose that we
encode a picture on the surface by affixing additional labels to edges.
Next let us raise the temperature until the surface disappears completely.
Then, we go back to the original temperature and a surface will form again.
The picture, however, will be scrambled, since the arrangement of
edges is completely reshuffled. One can view this as an instance of
spontaneous symmetry breaking.

Another version of this argument explains nicely the
intimate connection between space and Hamiltonian (or spacetime and Lagrangian).
Suppose that we extend the system by adding spins to the triangles
and let these spins interact by the usual nearest--neighbour Hamiltonian.
Assume furthermore that the energy scale of the spin field is small enough
to not affect the dynamics of the graphs.
At low enough temperatures, the graphs freeze in a surface state
and the system becomes effectively an Ising model on a fixed lattice.
Note, however, that it is the graph that determines which spin neighbours
which other spin and hence which spins couple to each other.
In this sense, the graph (or space) is part of the spin system's coupling
constants. By the same token, if space is a variable, the coupling constants
are variables as well.

\begin{acknowledgments}
I thank Leo Kadanoff and Seth Major for helpful discussions. I also thank Ky Le for assistance in installing the Ubigraph package.
This work was made possible by the facilities of the Shared Hierarchical
Academic Research Computing Network (SHARCNET:www.sharcnet.ca) and Compute/Calcul Canada.
Research at Perimeter Institute is supported by the Government of Canada through Industry Canada and by the Province of Ontario through the Ministry of Research \& Innovation.
\end{acknowledgments}

\end{document}